\documentclass[12pt,preprint]{aastex}

\slugcomment{Not to appear in Nonlearned J., 45}

\shorttitle{Kinematics of gas and stars in CNSFRs}
\shortauthors{H\"agele et al.}

\begin{document}


\title{Kinematics of gas and stars in circumnuclear star-forming regions of early type spirals}


\author{Guillermo F. H\"agele, \'Angeles
  I. D\'iaz and M\'onica V. Cardaci\altaffilmark{1}}
\affil{Dpto de F\'{\i}sica Te\'orica, C-XI, Universidad Aut\'onoma de
Madrid, 28049 Madrid, Spain}

\and

\author{Elena Terlevich\altaffilmark{2} and Roberto Terlevich\altaffilmark{2}}
\affil{INAOE, Tonantzintla, Apdo. Postal 51, 72000 Puebla, M\'exico}


\altaffiltext{1}{XMM-Newton Science Operations Center, ESAC, ESA, POB 78,
  E-28691 Villanueva de la Ca\~nada, Madrid, Spain.}
\altaffiltext{2}{Research Affiliate at IoA.}


\begin{abstract}
We present high resolution (R~20000) spectra in the blue and the far red of
cicumnuclear star-forming regions (CNSFRs) in three early type spirals
(NGC\,3351, NGC\,2903 and NGC\,3310) which have allowed the study of the
kinematics 
of stars and ionized gas in these structures and, for the first time, the
derivation of their dynamical masses for the first two. In some cases these
regions, about 100 to 150\,pc in size, are seen to be composed of several
individual star clusters with sizes between 1.5 and 4.9\,pc estimated from
Hubble Space Telescope (HST) images. The stellar dispersions have been
obtained from the Calcium triplet (CaT) lines at
$\lambda\lambda$\,8494,8542,8662\,\AA, while the gas velocity dispersions have
been measured by Gaussian fits to the H$\beta$ and
[OIII]\,$\lambda\lambda$\,5007\,\AA lines on the high dispersion
spectra. Values of the stellar velocity dispersions are between 30 and
68\,km/s. We apply the virial theorem to estimate dynamical masses of the
clusters, assuming that systems are gravitationally bounded and spherically
symmetric, and using previously measured sizes. The measured values of the 
stellar velocity dispersions yield dynamical masses of the order of 10$^7$ to
10$^8$ solar masses for the whole CNSFRs. Stellar and gas velocity dispersions
are found to differ by about 20 to 30\,km/s with the H$\beta$ emission lines
being narrower than both the stellar lines and the
[OIII]\,$\lambda\lambda$\,5007\,\AA lines. The twice ionized oxygen, on the
other hand, shows velocity dispersions comparable to those shown by stars, in
some cases, even larger. We have found indications of the presence of two
different kinematical components in the ionized gas of the regions. We have
mapped the velocity field in the central kiloparsec of the spiral galaxies
NGC\,3351 and NGC\,2903. For the first object the radial velocity curve shows
deviations from circular motions for the ionized hydrogen consistent with its
infall towards the central regions of the galaxy at a velocity of about
25\,km/s. For NGC\,3310 we present preliminary results of the velocity
dispersions for one of the two observed slit position angles, two CNSFRs and
the nucleus. 

\end{abstract}





\section{Observations and data reduction}

The high-resolution blue and far red spectra were acquired as part of an
observing run of three nights in 2000. They were simultaneously obtained with
ISIS, using the H2400B and R1200R gratings, respectively, on the 4.2m William
Herschell Telescope. In the blue arm the spectral range is 4780-5200\,\AA\ and
in the red arm it is 8360-8760\,\AA. The width of the slit is 1'' providing a
resolution of ~0.21\,\AA/pix and ~0.39\,\AA/pix, respectively. The spatial
resolutions are 0.4 and 0.39 arcsec/pix for the blue and red arms,
respectively. 
 
Three different slit position angles were chosen for NGC\,3351 to observe 5
CNSFRs and the nucleus \citep[see][]{hagele07}. In the case of NGC\,2903 two
different slit position angles allow us to study  4 CNSFRs and the
nucleus \citep[see][]{hagele08}. For NGC\,3310 we have observed in two slit
positions but we only 
present preliminary results for one of them. The images were processed and
analysed using  IRAF routines in the usual manner.  

In addition to the galaxy images, observations of 11 velocity template stars
were made to provide good stellar reference frames. They are late type giant
and supergiant stars which have strong CaT features \citep[see][]{diaz89}.

The different methods used to derivate the quantities presented here are
explained in detail in \cite{hagele07}.

\section{Summary and conclusions}

Stellar velocity dispersions are between 30 and 68\,km/s, about 25\,km/s
larger than those measured for the gas. However, the best Gaussian fits
involved two different components for the gas: a "broad component" with a
velocity dispersion  similar to that measured for the stars, and a "narrow
component" with a dispersion lower than the stellar one by about
30\,km/s. This last component seems to have a relatively constant value for
all the CNSFRs in each galaxy, with estimated values close to 25\,km/s for the
two gas emission lines. The velocities of the two components of the fits in
the CNSFRs of NGC\,3351 are the same within the observational errors, but in
the cases of NGC\,2903 and NGC\,3310 we find a shift between the narrow and
the broad component that vary between -10 and 35\,km/s in radial velocity. 

The dynamical masses estimated from the stellar velocity dispersion using the
virial theorem for the CNSFRs of NGC\,3351 are in the range between
4.9\,$\times\,10^6$ and 4.8\,$\times\,10^7$\,M$_\odot$, and is
3.5\,$\times\,10^7$ for its nuclear region inside the inner 11.3\,pc
\citep{hagele07}. In the 
case of NGC\,2903 the masses are in the range between 6.4\,$\times\,10^7$ and
2.1\,$\times\,10^8$\,M$_\odot$ for the CNSFRs and is 2.2\,$\times\,10^7$ for
its nuclear region inside the inner 3.8\,pc \citep{hagele08}. Masses derived
from the H$\beta$ velocity dispersions under the assumption of a single
component for the gas would have been underestimated by factors between
approximately 2 to 4.

Masses of the ionising stellar clusters of the CNSFRs have been derived from
their H$\alpha$ luminosities \citep{planesas97} under the assumption that the
regions are ionisation bound and without taking into account any photon
absorption by dust. Their values for NGC\,3351 are between 8.0\,$\times\,10^5$
and 2.5\,$\times\,10^6$ M$_\odot$ for the starforming regions, and is
6.0\,$\times\,10^5$  for the nucleus \citep{hagele07}. And for the regions of
NGC\,2903 they 
are between 3.3\,$\times\,10^6$ and 4.9\,$\times\,10^6$ M$_\odot$, and is
2.1\,$\times\,10^5$ for its nucleus \citep{hagele08}. These values are
comparable to that 
derived by \cite{gonzalez-delgado95} for the circumnuclear
region A in NGC\,7714. Therefore, the ratio of the ionising stellar population
to the total dynamical mass is between 0.01 and 0.16.  

Derived masses for the ionised gas vary between 7.0\,$\times\,10^3$ and
8.7\,$\times\,10^4$ M$_\odot$ for the CNSFRs of NGC\,3351 \citep{hagele07},
and is 2\,$\times\,10^3$ M$_\odot$ for the nucleus, and between
6.1\,$\times\,10^4$ and 1.3\,$\times\,10^5$ M$_\odot$ for the regions and is
2\,$\times\,10^3$ M$_\odot$ for the nucleus of NGC\,2903
\citep{hagele08}. These values are also comparable to that derived by
\cite{gonzalez-delgado95}.

For NGC\,3351, the rotation velocities derived for both stars and gas are in
reasonable agreement, although in some cases the gas shows a velocity slightly
different from that of the stars \citep{hagele07}. The rotation curve
corresponding to the 
position going through the centre of the galaxy shows maximum and minimum
values at the position of the circumnuclear ring, as observed in other
galaxies with CNSFRs \citep[][and reference therein]{diaz99}. The differences
in velocity between gas and stars can be interpreted 
as motions of the ionised hydrogen deviating from rotation and consistent with
a radial infall to the central regions of the galaxy. Our results are
consistent with those found by \cite{rubin75} and would yield
an infall velocity of about 25 km/s. 

On the other hand, the observed stellar and [OIII] velocities of NGC\,2903 are
in good agreement, while the H$\beta$ measurements show shifts similar to
those find between the narrow and the broad components \citep{hagele08}. This
different 
behavior can be due to that the position of the single Gaussian fits are
dominated by the broad component in the case of the [OIII] emission line while
in the case of the H$\beta$ are dominated by the narrow one. Again, the
rotation curve corresponding to the position going through the nucleus shows
maximum and minimum values at the positions of the circumnuclear regions.


\begin{thebibliography}{}
\bibitem[D\'iaz et al.(1989)]{diaz89} D\'iaz, A.~I., Terlevich, 
E., \& Terlevich, R.\ 1989, \mnras, 239, 325
\bibitem[D{\'{\i}}az et al.(1999)]{diaz99} D{\'{\i}}az, R., 
Carranza, G., Dottori, H., \& Goldes, G.\ 1999, \apj, 512, 623 
\bibitem[Gonzalez-Delgado et al.(1995)]{gonzalez-delgado95} 
Gonz\'alez-Delgado, R.~M., P\'erez, E., D\'iaz, A.~I., Garc\'ia-Vargas, M.~L., 
Terlevich, E., \& V\'ilchez, J.~M.\ 1995, \apj, 439, 604 
\bibitem[H{\"a}gele et al.(2007)]{hagele07} H{\"a}gele, G.~F., 
D{\'{\i}}az, {\'A}.~I., Cardaci, M.~V., Terlevich, E., \& Terlevich, R.\ 
2007, \mnras, 378, 163 
\bibitem[H{\"a}gele et al.(2008)]{hagele08} H{\"a}gele, G.~F., 
D{\'{\i}}az, {\'A}.~I., Cardaci, M.~V., Terlevich, E., \& Terlevich, R.\ 
2008, \mnras, in preparation 
\bibitem[Planesas et al.(1997)]{planesas97} Planesas, P., Colina, 
L., \& P\'erez-Olea, D.\ 1997, \aap, 325, 81 
\bibitem[Rubin et al.(1975)]{rubin75} Rubin, V.~C., Peterson, 
C.~J., \& Ford, W.~K., Jr.\ 1975, \apj, 199, 39 
\end{thebibliography}
\end{document}